%% file: cipanp2015_strauch.tex
\documentclass[12pt]{article}
\usepackage{graphicx}
\usepackage{amssymb,amsmath}
\usepackage{hyperref}
\usepackage{float}
\usepackage[utf8]{inputenc}

\newcommand\pubnumber{CIPANP2015-Strauch}
\newcommand\pubdate{\today}

\def\napoli{Department of Physics \& Astronomy\\
University of South Carolina, Columbia, SC 20208, USA}
\def\support{\footnote{Supported in parts by the U.S. National Science Foundation: NSF PHY-1205782.}}

\def\Title#1{\begin{center} {\Large #1 } \end{center}}
\def\Author#1{\begin{center}{ \sc #1} \end{center}}
\def\Address#1{\begin{center}{ \it #1} \end{center}}

\newcommand\pubblock{\rightline{\begin{tabular}{l} \pubnumber\\
         \pubdate  \end{tabular}}}
\newenvironment{Abstract}{\begin{quotation}  }{\end{quotation}}
\newenvironment{Presented}{\begin{quotation} \begin{center} 
             PRESENTED AT\end{center}\bigskip 
      \begin{center}\begin{large}}{\end{large}\end{center} \end{quotation}}

\input econfmacros.tex

\newcommand{\polpipn}{$\vec\gamma \vec p\to\pi^+n$ {}}

\begin{document}
\begin{titlepage}
\pubblock

\vfill
\Title{Baryon spectroscopy with polarization observables from CLAS}
\vfill
\Author{Steffen Strauch\support for the CLAS Collaboration}
\Address{\napoli}
\vfill
\begin{Abstract}
  Meson photoproduction is an important tool in the study of baryon
  resonances. The spectrum of broad and overlapping nucleon
  excitations can be greatly clarified by use of polarization
  observables. The N* program at Jefferson Lab with the CEBAF Large
  Acceptance Spectrometer (CLAS) includes experimental studies with
  linearly and circularly polarized tagged photon beams,
  longitudinally and transversely polarized nucleon targets, and
  recoil polarizations. An overview of these experimental studies and
  recent results will be given.
\end{Abstract}
\vfill
\begin{Presented}
Conference on the Intersections of
Particle and Nuclear Physics (CIPANP)\\ Vail, CO, USA, May 19--24, 2015
\end{Presented}
\vfill
\end{titlepage}
\def\thefootnote{\fnsymbol{footnote}}
\setcounter{footnote}{0}
%


\section{Introduction}

The nucleon is a composite object of quarks and gluons.  The 
spectrum of its excited states reveals valuable information about the
fundamental theory of strong interaction, quantum chromodynamics
(QCD).  The spectrum also gives information about the relevant degrees
of freedom of the bound system.  Various models may differ in their
underlying degrees of freedom, e.g. symmetric quark models,
quark-diquark models, models which include gluonic excitations, and
meson-baryon models. All symmetric quark models predict an
overabundance of excited states relative to what has been observed
until now, especially in $\pi N$ scattering.  Also recent lattice QCD
calculations \cite{Edwards:2011jj} yield a large number of not yet
discovered nucleon resonances.

The large widths of the overlapping resonances make them difficult to
detect and cross section data alone are insufficient to isolate
resonance contributions and are insufficient to fully constrain
partial-wave analyses.  Polarization observables are crucial in these
analyses.  These observables include singe- and double-polarization
observables with combinations of polarized beam, target, or the
polarization of the recoiling baryon~\cite{Barker:1975bp}.
Eventually, a complete set of certain polarization observables is
necessary to unambiguously determine the amplitudes of the reaction.
In the photoproduction of pseudoscalar mesons a formally complete experiment
requires the measurement of at least eight appropriately chosen
observables at each energy and angle \cite{Chiang:1996em}.  In the
photoproduction of two mesons, more observables are needed
\cite{Roberts:2005mn}.

In the following, examples of recent photoproduction measurements of
polarization observables from the CLAS Collaboration are presented.
These studies include single pseudoscalar-meson, vector-meson,
double-pion, and hyperon photoproduction off the proton and
(quasi-free) off the bound neutron.  The experiments were performed at
the Thomas Jefferson National Accelerator Facility (JLab).  The
incident photon beams were energy tagged in the Hall-B Photon Tagger
and \cite{Sober:2000we} either unpolarized, circularly, or linearly
polarized. The photon beam irradiated the production target.
Unpolarized liquid hydrogen or deuterium or the newly developed
polarized frozen-spin (FROST) \cite{Keith201227} or HDice targets
\cite{Sandorfi:2013gra,PR06101} have been used in these experiments.
Final-state particles were detected in the CEBAF Large Acceptance
Spectrometer (CLAS) \cite{Mecking:2003zu}. The parity-violating weak
decay of hyperons allows the determination of the hyperon polarization
by measuring the decay-proton angular distributions.

\section{Unpolarized Target}

Finely binned beam asymmetries, $\Sigma$, have been measured for the
reactions $\gamma p \to p\pi^0$ and $\gamma p \to n\pi^+$ with
linearly polarized photons and energies from 1.10 to 1.86~GeV
\cite{Dugger:2013crn}.  The observables were extracted from angular
distributions of the polarized yields with respect to the polarization
direction of the photon beam.  The data strongly constrained earlier
partial-wave analyses.  Resonance couplings have been extracted in
fits made with the SAID analysis.  The largest change from previous
fits was found to occur for the 'well known' $\Delta(1700)3/2^-$ and
$\Delta(1905)5/2^+$ resonances \cite{Dugger:2013crn}.

The CLAS Collaboration has measured the $\Lambda$ recoil polarization,
$P$ \cite{McNabb:2003nf,McCracken:2009ra}, as well as the beam-recoil
observables, $C_x$ and $C_z$ \cite{Bradford:2006ba}, in the reactions
$\gamma p \to K^+\Lambda$ and $\gamma p \to K^+\Sigma^0$.  
These data were instrumental in a coupled-channel analyses of the
Bonn-Gatchina group \cite{Sarantsev:2005tg,Nikonov:2007br}.  In
particular, the analysis found further evidence for the, at the time,
poorly known $N(1900)3/2^+$ resonance.  The $N(1900)3/2^+$ is a
resonance which is predicted by symmetric three-quark models, but is
not expected to exist in earlier quark-diquark models.

\begin{figure}[htb]
\centering
\includegraphics[width=\textwidth]{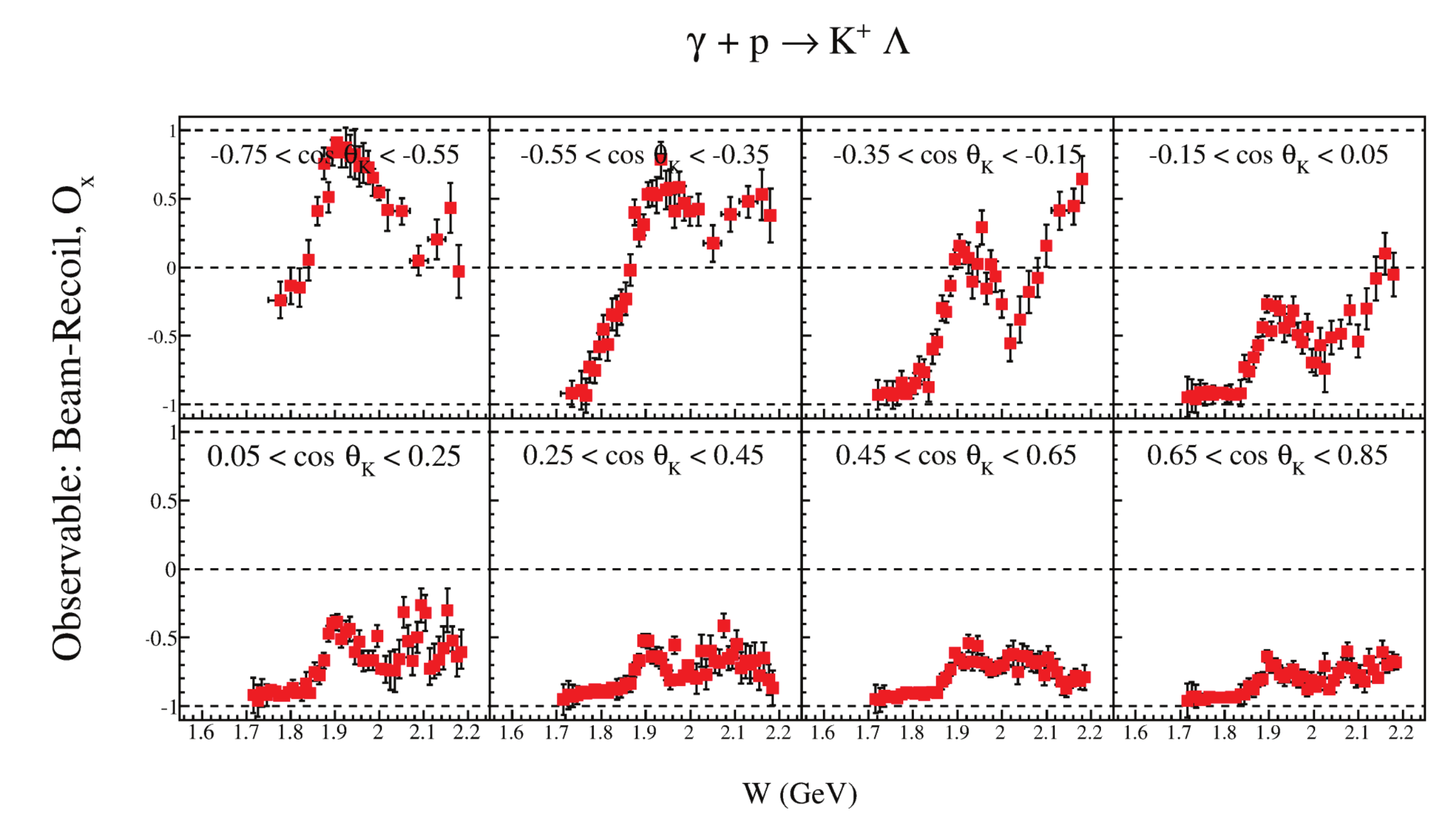}
\caption{Preliminary results of CLAS data for the double-polarization
  observable $O_x$ in the $\gamma p \to K^+ \Lambda$ reaction.  Figure
from D.~Ireland (University of Glasgow).}
\label{fig:cipanp2015_strauch_fig2}
\end{figure}

These hyperon photoproduction studies are being extended in a new
analysis of data with linearly polarized photons off a proton target
up to $W \approx 2.2$~GeV.  Together with the recoil polarization of
the hyperon, this gives access to five polarization observables:
$\Sigma$, $P$, $T$, $O_x$, and $O_z$.  Energy distributions of the
preliminary results of the beam-recoil polarization observable $O_x$
are shown in Fig.~\ref{fig:cipanp2015_strauch_fig2} for the
$\gamma p \to K^+ \Lambda$ reaction.

\section{Polarized Proton Target (FROST)}

The CLAS frozen-spin target program combines a series of meson
photoproduction experiments
\cite{PR02112,PR03105,PR04102,PR05012,PR06013} to measure single- and
double-polarization observables with polarized photons off
longitudinally or transversely polarized proton (butanol) targets.

\begin{figure}[htb]
\centering
\includegraphics[width=\textwidth]{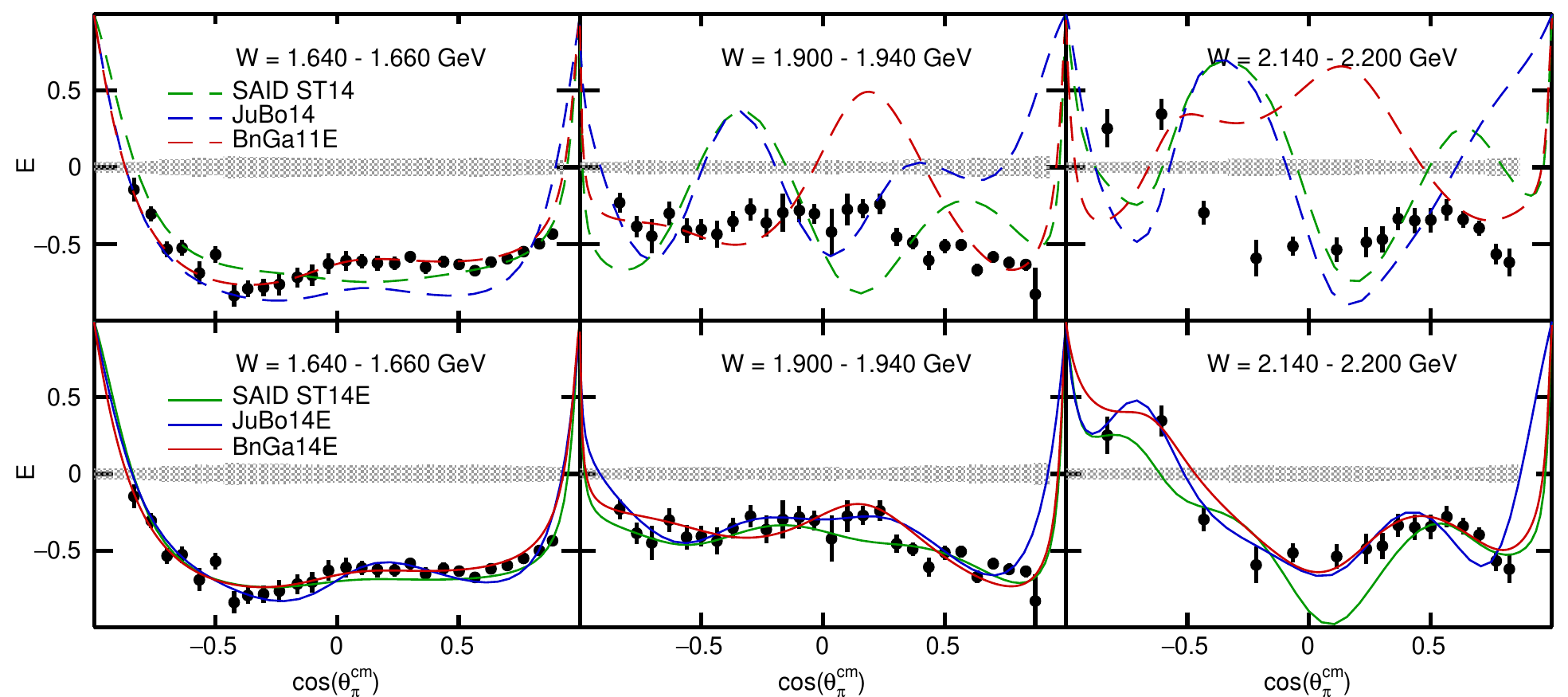}
\caption{Observable $E$ in the \polpipn reaction for selected
  center-of-mass energy bins.  Systematic uncertainties are indicated
  as shaded bands.  The curves in the upper panels are results from
  the SAID ST14 \cite{Workman:2012jf}, J\"ulich14
  \cite{Ronchen:2014cna}, and BnGa11E \cite{Anisovich:2011fc}
  analyses.  The curves in the lower panels are results from updated
  analyses including the present $E$ data. Figure from
  Ref.~\cite{Strauch:2015zob}.}
\label{fig:cipanp2015_strauch_fig1}
\end{figure}

First results have been published for the double-polarization
observable $E$ in the $\vec \gamma \vec p \to \pi^+n$ with circularly
polarized tagged-photon beam, with energies from 0.35 to 2.37~GeV, off
longitudinally polarized protons \cite{Strauch:2015zob}.  Previous
partial-wave analyses describe the new data at low photon energies
reasonably well, at high energies, however, significant deviations are
observed; see Fig.~\ref{fig:cipanp2015_strauch_fig1}.  The data have
been included in new multipole analyses resulting in updated nucleon
resonance parameters.  One particularly interesting result is
strengthened evidence for the poorly known
$\Delta (2200) {7 \over 2}^-$ resonance in improving the Bonn-Gatchina
fit at the highest energies~\cite{Anisovich:2015gia}.  Its mass is
significantly higher than the mass of its parity partner
$\Delta(1950)7/2^+$ which is the lowest-mass $\Delta^*$ resonance with
spin-parity $ J^P=7/2^+$.
FROST data from other single-pion photoproduction channels are under
analysis by Arizona State University, U. of South Carolina, and U. of
Edinburgh groups.

\begin{figure}[htb]
\centering
\includegraphics[width=0.7\textwidth]{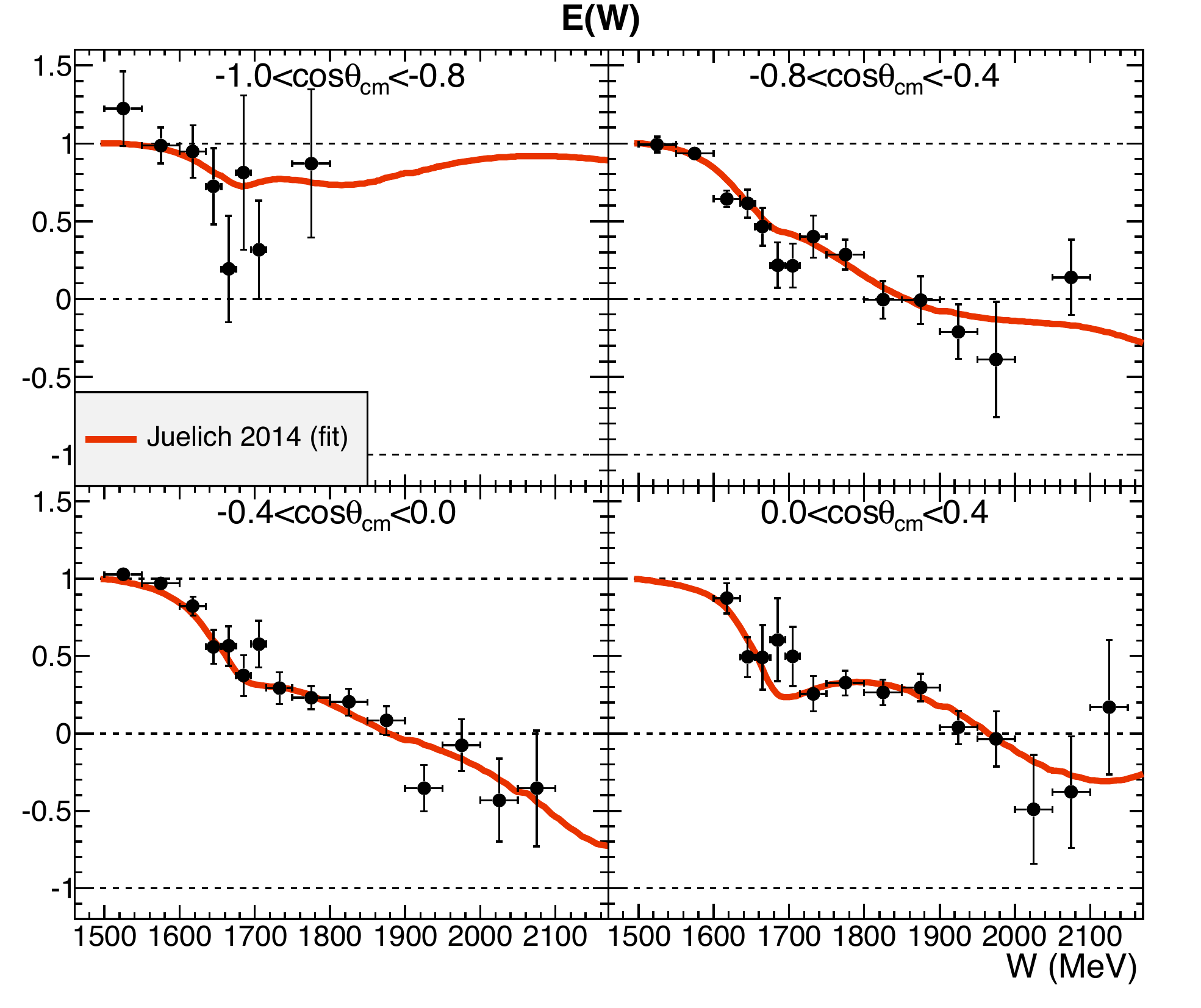}
\caption{Observable $E$ in the$\gamma p \to \eta p$ as a function of
  $W$ for various angular bins. Figure from Ref.~\cite{Senderovich:2015lek}.}
\label{fig:cipanp2015_strauch_fig3}
\end{figure}

The polarization observable $E$ has also been measured up to
$W = 2.15$~GeV in the $\gamma p \to \eta p$ reaction
\cite{Senderovich:2015lek}; this reaction selects isospin-$1/2$
resonances.  Figure \ref{fig:cipanp2015_strauch_fig3} shows the data
and a fit of the J\"ulich model.  The fit describes the data quite
well without the need for an additional narrow resonance near
1.68~GeV, which was previously suggested; see
\cite{Senderovich:2015lek}.

\begin{figure}[H]
\centering
\includegraphics[width=0.8\textwidth]{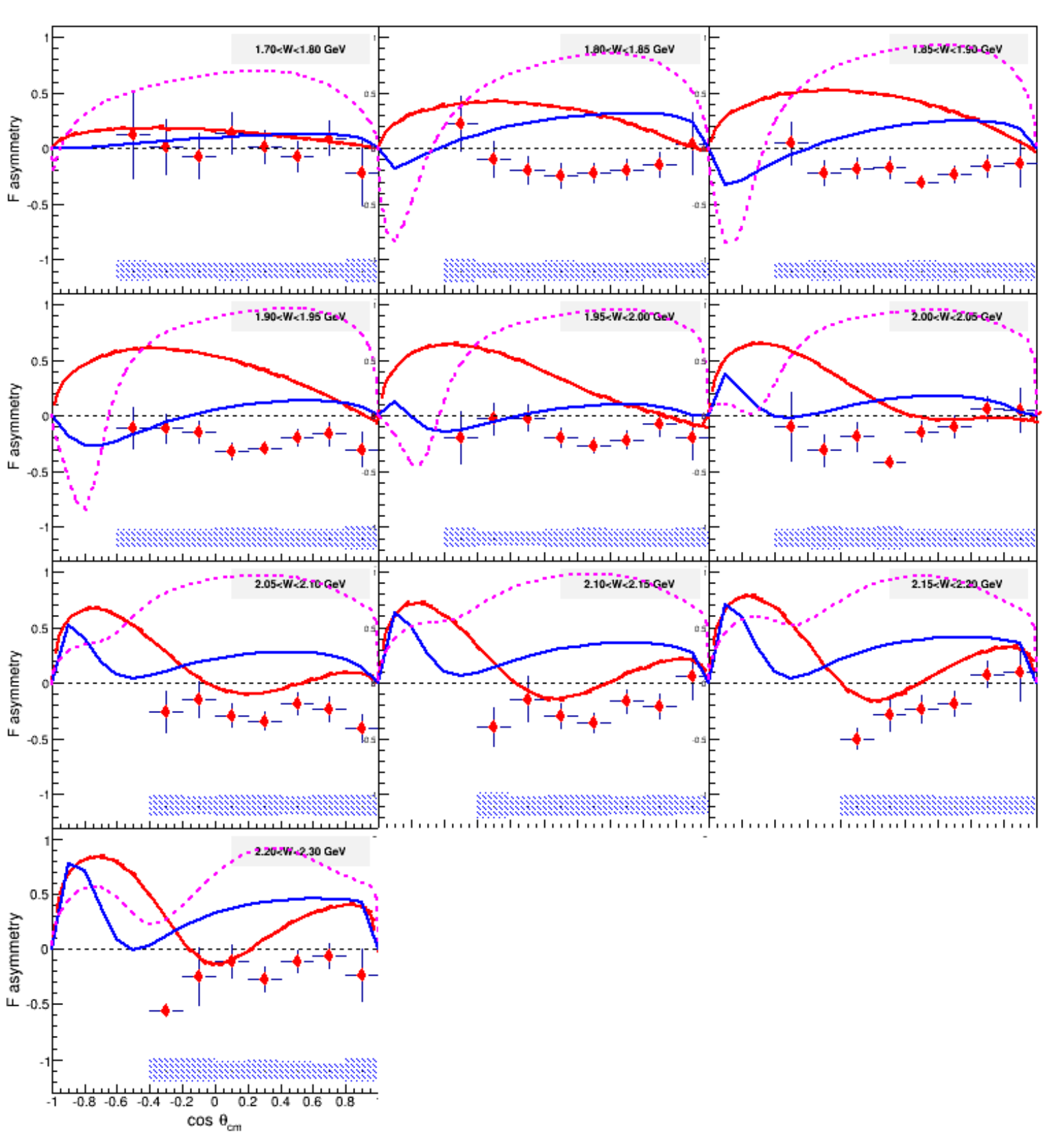}
\caption{Preliminary angular distributions of the polarization
  observable $F$ in the $\gamma p \to K^+\Sigma^0$ reaction for
  various $W$ bins.  Model curves are from the RPR-Ghent \cite{WalfordRPR} (red),
  KAON-MAID \cite{WalfordKM}  (blue), and Bonn-Gatchina \cite{Anisovich:2011fc} (magenta) models.  Figure and
  analysis from N.~Walford (U.~Basel).}
\label{fig:cipanp2015_strauch_fig4}
\end{figure}

Hyperon photoproduction reactions are also studied with data from
FROST.  Figure \ref{fig:cipanp2015_strauch_fig4} shows preliminary
results of the double polarization observable $F$, which can be
measured with circularly polarized photons off transversely polarized
protons.  The data cover energies between $W = 1.7$~GeV and
$W = 2.3$~GeV.  The data are compared with previous model calculations
from the RPR-Ghent \cite{WalfordRPR}, KAON-MAID \cite{WalfordKM}, and
Bonn-Gatchina \cite{Anisovich:2011fc}  models; none of the
models describe the data well, showing the new constraints the data
will provide.

\begin{figure}[htb]
\centering
\includegraphics[width=\textwidth]{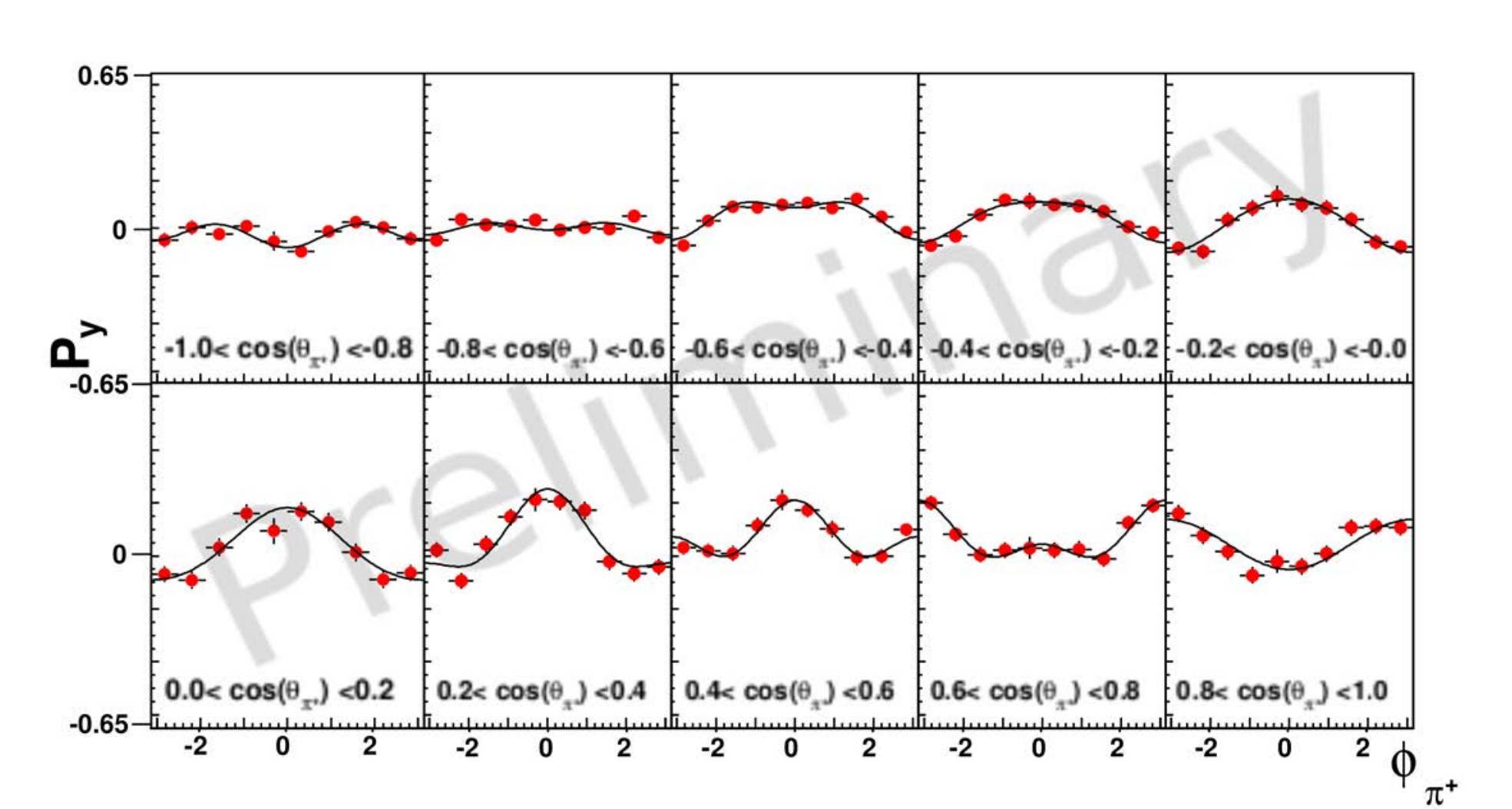}
\caption{Preliminary results of the target polarization observable
  $P_y$ in the $\vec\gamma p \to p \pi^+\pi^-$ reaction for
  $1.1~\text{GeV} < E_\gamma < 1.2~\text{GeV}$. The angles $\theta_{\pi^+}$
  and $\phi_{\pi^+}$ are defined as the polar and azimuthal angles in
  the rest frame of the $\pi^+\pi^-$ system with the $z$ direction
  along the total momentum of the $\pi^+\pi^-$ system.  The data are
  fitted with a low-order Fourier series (black curve). Figure and
  analysis by P.~Roy (FSU).}
\label{fig:cipanp2015_strauch_fig5}
\end{figure}

Many nucleon resonances in the mass region above 1.6 GeV decay
predominantly through either $\pi\Delta$ or $\rho N$ intermediate
states into $\pi\pi N$ final states. This makes double-pion
photoproduction an important tool in the investigation of the
structure of the nucleon.  The CLAS Collaboration was first to study
the beam-helicity asymmetry $I^\odot$ for the two-pion-photoproduction
reaction.  The measurement covered energies between $W = 1.35$ and
2.30 GeV \cite{Strauch:2005cs}.  With the development of the polarized
FROST target many more polarization observables became accessable; see
\cite{Roberts:2005mn} for an overview of polarization observables in
two-pion photoproduction reactions.  The University of South Carolina
and Florida State University groups work on the extraction of twelve
different polarization observables from polarized-target
$p \pi^+\pi^-$ data.  As example, the preliminary result for the
observable $P_y$ is shown in Fig.~\ref{fig:cipanp2015_strauch_fig5}.
This observable is accessable in a measurement with unpolarized
photons off transversally polarized protons.  The data will strongly
constrain coupled-channel analyses.

\section{Polarized Neutron Target (HDice)}

So far, the majority of data have been taken off proton targets.  The
new data from the FROST program will expand the data base over a large
range of energy with many observables for polarized proton reactions.
In constrast, data off neutrons are extremely sparse.  However,
measurements with both proton and neutron targets are needed to
completely specify the amplitude of the reaction.

The CLAS collaboration has taken production data with circularly and
linearly polarized photons off a polarized solid deuterium-hydride
target (HDice) \cite{Sandorfi:2013gra,PR06101} up to center-of-mass
energies of $W \approx 2.3$~GeV.  The run conditions
were optimized for polarized neutron reactions.  The ongoing analyses
of this run include single- and double pion photoproduction and
hyperon photoproduction off the bound neutron.

\section{Conclusion}

During the past years the knowledge of the baryon spectrum has largely
increased, to a large extend due to photoproduction data from JLab,
ELSA, GRAAL, and LEPS.

New polarized photoproduction data from CLAS off polarized and
unpolarized, proton and neutron targets are under analysis and become
available.  They will contribute to complete or nearly complete
experiments.  They challenge previously not sufficiently constrained
models. It is very likely that they will have a tremendous impact on
the understanding of baryon resonances and may provide evidence for
new states found in coupled-channel analyses.


\bibliographystyle{cipanp2015_style}
\bibliography{frost}

\end{document}

%% file: econfmacros.tex
\def\beq{\begin{equation}}
\def\eeq#1{\label{#1}\end{equation}}
\def\eeqn{\end{equation}}

\def\beqa{\begin{eqnarray}}
\def\eeqa#1{\label{#1}\end{eqnarray}}
\def\eeqan{\end{eqnarray}}

\let\bar=\overbar

\def\Dslash{\not{\hbox{\kern-4pt $D$}}}
\def\dslash{\not{\hbox{\kern-2pt $\del$}}}

\def\msb{{\bar{\ssstyle M \kern -1pt S}}}




%% file: cipanp2015_strauch.bbl
\begin{thebibliography}{10}
\providecommand{\url}[1]{\texttt{#1}}
\providecommand{\urlprefix}{URL }
\providecommand{\eprint}[2][]{\url{#2}}

\bibitem{Edwards:2011jj}
R.~G. Edwards, J.~J. Dudek, D.~G. Richards, and S.~J. Wallace, Phys.~Rev.
  \textbf{D84}, 074508 (2011), \eprint{1104.5152}.

\bibitem{Barker:1975bp}
I.~S. Barker, A.~Donnachie, and J.~K. Storrow, Nucl.~Phys. \textbf{B95}, 347
  (1975).

\bibitem{Chiang:1996em}
W.-T. Chiang and F.~Tabakin, Phys.~Rev. \textbf{C55}, 2054 (1997).

\bibitem{Roberts:2005mn}
W.~Roberts and T.~Oed, Phys.Rev. \textbf{C71}, 055201 (2005),
  \eprint{nucl-th/0410012}.

\bibitem{Sober:2000we}
D.~Sober \emph{et~al.}, Nucl.~Instrum.~Meth.~A \textbf{440}, 263 (2000).

\bibitem{Keith201227}
C.~Keith, J.~Brock, C.~Carlin, S.~Comer, D.~Kashy, J.~McAndrew, D.~Meekins,
  E.~Pasyuk, J.~Pierce, and M.~Seely, Nucl.~Instrum.~Meth.~A \textbf{684}, 27
  (2012).

\bibitem{Sandorfi:2013gra}
A.~M. Sandorfi, J. Phys. Conf. Ser. \textbf{424}, 012001 (2013).

\bibitem{PR06101}
{Jefferson Lab Experiment E06-101, ``N* Resonances in Pseudoscalar-meson
  photo-production from Polarized Neutrons in $\vec H \cdot \vec D$ and a
  complete determination of the $\gamma n \to K^0\Lambda$ amplitude'', F.~Klein
  and A.M.~Sandorfi, spkespersons}.

\bibitem{Mecking:2003zu}
B.~A. Mecking \emph{et~al.}, Nucl.~Instrum.~Meth. \textbf{A503}, 513 (2003).

\bibitem{Dugger:2013crn}
M.~Dugger \emph{et~al.}, Phys.~Rev. C \textbf{88}, 065203 (2013),
  \eprint{1308.4028}.

\bibitem{McNabb:2003nf}
J.~W.~C. McNabb \emph{et~al.}, Phys. Rev. \textbf{C69}, 042201 (2004),
  \eprint{nucl-ex/0305028}.

\bibitem{McCracken:2009ra}
M.~E. McCracken \emph{et~al.}, Phys. Rev. \textbf{C81}, 025201 (2010),
  \eprint{0912.4274}.

\bibitem{Bradford:2006ba}
R.~K. Bradford \emph{et~al.}, Phys. Rev. \textbf{C75}, 035205 (2007),
  \eprint{nucl-ex/0611034}.

\bibitem{Sarantsev:2005tg}
A.~V. Sarantsev, V.~A. Nikonov, A.~V. Anisovich, E.~Klempt, and U.~Thoma, Eur.
  Phys. J. \textbf{A25}, 441 (2005), \eprint{hep-ex/0506011}.

\bibitem{Nikonov:2007br}
V.~A. Nikonov, A.~V. Anisovich, E.~Klempt, A.~V. Sarantsev, and U.~Thoma, Phys.
  Lett. \textbf{B662}, 245 (2008), \eprint{0707.3600}.

\bibitem{PR02112}
Jefferson Lab Experiment E02-112, ``Search for Missing Nucleon Resonances in
  Hyperon Photoproduction'', P.~Eugenio, F.~Klein, and L.~Todor, spokespersons.

\bibitem{PR03105}
Jefferson Lab Experiment E03-105, ``Pion Photoproduction from a Polarized
  Target'', N.~Benmouna, W.~Briscoe, G.~O'Rielly, I.~Strakovsky, S.~Strauch,
  spokespersons.

\bibitem{PR04102}
Jefferson Lab experiment E04-102, ``Helicity Structure of Pion
  Photoproduction'', D.~Crabb, M.~Khandaker, and D.~Sober, spokespersons.

\bibitem{PR05012}
Jefferson Lab experiment E05-012, ``Measurement of polarization observables in
  $\eta$-photoproduction with CLAS'', M.~Dugger and E.~Pasyuk, spokespersons.

\bibitem{PR06013}
Jefferson Lab Experiment E06-013, ``Measurement of $\pi^+\pi^-$ Photoproduction
  in Double-Polarization Experiments using CLAS'', M.~Bellis, V.~Cred\'{e},
  S.~Strauch, spokespersons.

\bibitem{Strauch:2015zob}
S.~Strauch \emph{et~al.}, Phys. Lett. \textbf{B750}, 53 (2015),
  \eprint{1503.05163}.

\bibitem{Anisovich:2015gia}
A.~V. Anisovich, V.~Burkert, E.~Klempt, V.~A. Nikonov, E.~Pasyuk, A.~V.
  Sarantsev, S.~Strauch, and U.~Thoma (2015), \eprint{nucl-ex/1503.05774}.

\bibitem{Workman:2012jf}
R.~L. Workman, M.~W. Paris, W.~J. Briscoe, and I.~I. Strakovsky, Phys.~Rev.
  \textbf{C86}, 015202 (2012).

\bibitem{Ronchen:2014cna}
D.~Rönchen, M.~Döring, F.~Huang, H.~Haberzettl, J.~Haidenbauer, C.~Hanhart,
  S.~Krewald, U.~G. Meißner, and K.~Nakayama, Eur. Phys. J. \textbf{A50}, 101
  (2014), \eprint{1401.0634}.

\bibitem{Anisovich:2011fc}
A.~V. Anisovich, R.~Beck, E.~Klempt, V.~A. Nikonov, A.~V. Sarantsev, and
  U.~Thoma, Eur. Phys. J. \textbf{A48}, 15 (2012), \eprint{1112.4937}.

\bibitem{Senderovich:2015lek}
I.~Senderovich \emph{et~al.} (2015), \eprint{nucl-ex/1507.00325}.

\bibitem{WalfordRPR}
T. Corthals, J. Ryckebusch, and T. Van Cauteren, Phys. Rev. {\bf C73} (2006)
  045207 and T.~Vrancx, personal communication.

\bibitem{WalfordKM}
F. X. Lee, T. Mart, C. Bennhold, H. Haberzettl, and L. E. Wright, Nucl. Phys.
  {\bf A695}, 237 (2001), ``Kaon-MAID". Available from
  http://www.kph.uni-mainz.de/MAID/kaon/kaonmaid.html.

\bibitem{Strauch:2005cs}
S.~Strauch \emph{et~al.}, Phys. Rev. Lett. \textbf{95}, 162003 (2005),
  \eprint{hep-ex/0508002}.

\end{thebibliography}
